\documentclass[aps,twocolumn,showpacs,superscriptaddress,nofootinbib]{revtex4-2}

\usepackage{graphicx}
\usepackage{latexsym}

\usepackage{color}

\begin{document}
\title{
Projection of strong coupling interaction with thermal bath in a polymer
}

\author{Takuya Saito}
\email[Electric mail:]{tsaito@phys.aoyama.ac.jp}
\affiliation{Department of physical sciences, Aoyama Gakuin University, Chuo-ku, Sagamihara 252-5258, Japan}

\def\Vec#1{\mbox{\boldmath $#1$}}
\def\degC{\kern-.2em\r{}\kern-.3em C}

\def\SimIneA{\hspace{0.3em}\raisebox{0.4ex}{$<$}\hspace{-0.75em}\raisebox{-.7ex}{$\sim$}\hspace{0.3em}} 

\def\SimIneB{\hspace{0.3em}\raisebox{0.4ex}{$>$}\hspace{-0.75em}\raisebox{-.7ex}{$\sim$}\hspace{0.3em}}

\date{\today}

\begin{abstract}
We investigate modifications of a stochastic polymer picture through a shift in the boundary between the system and an external environment.
A conventional bead-and-spring model serving as the coarse-graining model is given by the Langevin equation for all the monomers subject to white noise.
However, stochastic motion for only a tagged monomer is observed to occur in the presence of colored noise. 
The qualitative change in the observations arises from the boundary shift decided by the observer. 
The Langevin dynamics analyses interpret the colored noise as the emergence of the polymeric elastic force, resulting in additional heat in the tagged monomer observation.
Being distinguished from coarse-graining based on scale separation, the projection of comparable internal degrees of freedom is also discussed in light of the fluctuation theorem and the stochastic polymer thermodynamics.
\end{abstract}


\def\degC{\kern-.2em\r{}\kern-.3em C}

\newcommand{\gsim}{\hspace{0.3em}\raisebox{0.5ex}{$>$}\hspace{-0.75em}\raisebox{-.7ex}{$\sim$}\hspace{0.3em}} 
\newcommand{\lsim}{\hspace{0.3em}\raisebox{0.5ex}{$<$}\hspace{-0.75em}\raisebox{-.7ex}{$\sim$}\hspace{0.3em}}

\maketitle

\section{Introduction}

Identification of thermodynamic quantities at a small scale is intrinsically involved in a distinct notion from the macroscopic framework~\cite{SekimotoBook,PRE_Gelin_Thoss_2009,PRL_Seifert_2016,PRE_Talkner_Hanggi_2016,PRX_Jarzynski_2017}. 
Remarkable differences arise from surface effects and fluctuations, both of which are eliminated at the thermodynamic limit, where the surface-to-bulk volume ratio approaches zero. Also, the law of large numbers provides a guide to view thermodynamic quantities as being definite values.
However, at the small scale, the interaction energy with a thermal bath could be comparable to the system energy itself. This mode of interaction is referred to as strong coupling and has been developed to be incorporated into stochastic thermodynamics~\cite{PRE_Gelin_Thoss_2009,PRL_Seifert_2016,PRE_Talkner_Hanggi_2016,PRX_Jarzynski_2017} together with the concept of entropy at the small scale.

A polymer consisting of numerous monomers is a strong coupling system. 
Indeed, a bead-and-spring model well known as the coarse-grained picture has only an effective Hamiltonian that does not rely on details of the specific structures because of relaxation due to interaction with a thermal bath~\cite{deGennesBook,Doi_Edwards,Khoklov_Grosberg}.
The polymer system also poses an interesting issue for projection methods~\cite{StatisticalPhysics_II,PTP_Mori_1965}.
If all the monomers are observed, they receive direct thermal agitation, which is commonly modeled by white noise in a viscous solution~\cite{deGennesBook,Doi_Edwards,Khoklov_Grosberg}.
However, if only a monomer is traced and the others are projected, the other monomers are only in a position of indirect interaction with the thermal bath, where the traced monomer is subject to colored noise~\cite{JChemPhys_Schiessel_Oshanin_Blumen_1995,PRE_Lizana_Barkai_Lomholt_2010,JStatMech_Panja_2010,PRE_Sakaue_2013,PRE_Saito_2015,StatisticalPhysics_II}.
The change in observed noise is noteworthy from the perspective of stochastic energetics~\cite{SekimotoBook,EPL_Harada_2005,JStatMech_Ohkuma_Ohta_2007} because the issue of how to identify the noise characteristics is closely related to a definition of heat made to satisfy first law of the thermodynamics. 
Another noteworthy point is that projections based on scale separation are distinguished from the elimination of degrees of freedom of the other monomers with a comparable spatiotemporal resolution, which corresponds to a recognition shift of the surface boundary between the system and the external environment.
We are then naturally led to the idea that the projection inherent in the polymer modifies the stochastic physical picture because of the boundary shift.
However, this concept has not been fully discussed.

This study focuses on how observation of a polymer system strongly coupled with a thermal bath is dependent on the projections of the degrees of freedom, apart from the scale separation.
From this perspective, section~\ref{Rouse} analyzes the Rouse model in a rigorous manner using Langevin dynamics~\cite{EPJB_Speck_Seifert_2005,PRE_Dhar_2005,PRE_Sharma_Cherayil_2011,PRE_Sakaue_2012}, where two observation methods-tracing a single monomer and tracing all the monomers-are investigated in view of heat under a stochastic energetics approach~\cite{SekimotoBook}. 
Section~\ref{GEq} describes the development of the general model for a polymer in a viscous solution by incorporating nonlocal effects of self-avoidance and hydrodynamic interactions.
Section~\ref{ThermoDy} identifies heat deduced from the fluctuation theorem.
The final arguments address stochastic thermodynamics based on a solvated ensemble~\cite{PRX_Jarzynski_2017}.
In section ~\ref{conclusion}, we summarize the study.

\section{Rouse polymer}
\label{Rouse}

We begin with the Rouse polymer model in Langevin dynamics.
A linear polymer chain consists of $N$ monomers of size $a$, which are labeled from one end, e.g., $x_n(t)$ denotes $n$-th monomer's position along a forced direction.
The one end ($N$-th monomer) begins to be pulled at $t=0$ by force $f$.
Unless otherwise noted, the $N$-th forced monomer is referred to as the tagged monomer, whose subscript situationally drops as $x(t)\equiv x_N(t)$.
Throughout the article, we consider near-equilibrium dynamics, where the external force is so weak, $f<k_BT/R_0\, (\simeq k_BT/(aN^\nu))$, that the polymer can qualitatively retain its equilibrium shape.
Note that $R_0=aN^{\nu}$ is an equilibrium coil size, and the Rouse model takes on $\nu=1/2$\footnote{The Flory exponents $\nu$ will be mentioned again after eq.~(\ref{mode_spring_friction})}.

\subsection{Langevin equation}

{\it --- Langevin equation ---}
The equation of motion is written with overdamped Langevin dynamics as~\cite{deGennesBook,Doi_Edwards,Khoklov_Grosberg,PRE_Sakaue_2012,JChemPhys_Schiessel_Oshanin_Blumen_1995,PRE_Lizana_Barkai_Lomholt_2010,JStatMech_Panja_2010,PRE_Saito_2015}:
\begin{eqnarray}
\gamma \frac{\partial x_n(t)}{\partial t}
=
-\frac{\partial {\cal H}^{(A)}(\{ x_n\})}{\partial x_n}
+f_n(t)+\zeta_n(t).
\label{Rouse_em}
\end{eqnarray}
The left side of this equation represents frictional force in a Newtonian fluid with frictional coefficient $\gamma$.
On the right side, the respective terms are, from left to right, the conservative force produced by the effective Hamiltonian ${\cal H}^{(A)}(\{ x_n\})$, the time-dependent external force $f_n(t)$, and thermal agitation $\zeta_n(t)$. 
Step force $f_{n}(t)=f\delta_{nN}\Theta (t)$ is applied to the tagged $N$-th monomer, and $\Theta (t)$ denotes the Heaviside step function.
The effective Hamiltonian is responsible for the internal interaction between the monomers to form chained configurations.
The effective Hamiltonian for the Rouse model is written as the sum of harmonic potentials between the adjoining monomers:
\begin{eqnarray}
{\cal H}^{(A)}(\{ x_n\}) &=& \sum_{n=-1}^N \frac{k}{2}(x_{n+1}-x_{n})^2
\label{Rouse_Hamiltonian}
\end{eqnarray}
with spring constant $k$.
Technically, $x_{-1}\equiv x_0$ and $x_{N+1}\equiv x_N$ are assumed to represent a boundary condition~\cite{deGennesBook}.
The random force of the thermal agitation satisfies the zero mean and the Gaussian-distributed noise with covariance:
\begin{eqnarray}
\left< \zeta_n(t)\zeta_{n'}(t') \right>=2k_BT \gamma \delta_{nn'} \delta (t-t').
\end{eqnarray}
Notably, eq.~(\ref{Rouse_em}) is the time evolution equation on the coordinate of all the monomers, whereas the generalized Langevin equation (GLE)~(\ref{GLE_mu}) is that on the coordinate of only the tagged monomer.
To solve eq.~(\ref{Rouse_em}), the normal mode $X_q(t)$ is used with the following transform:\footnote{
As in the literature~\cite{Doi_Edwards}, a long chain may technically replace eqs.~(\ref{variable_normal}) with
\begin{eqnarray}
X_q(t) \equiv \int_{0}^N dn\, x_n(t) h_{q,n}, \quad x_n(t)=\sum_{q=0}^N X_q(t)h_{q,n}^\dagger.
\end{eqnarray}
}
\begin{eqnarray}
X_q(t) \equiv \sum_{n=0}^N  x_n(t) h_{q,n}, \quad x_n(t)=\sum_{q=0}^N X_q(t)h_{q,n}^\dagger,
\label{variable_normal}
\end{eqnarray}
The kernels are defined as
\begin{eqnarray}
h_{q,n}\equiv \frac{1}{N}\cos{\left( \frac{qn \pi}{N} \right)},\quad h_{q,n}^\dagger = \frac{1}{c_q} \cos{\left( \frac{qn\pi}{N}\right)},
\label{kernel_h}
\end{eqnarray}
where $q$ denotes the mode indices; we set $c_q=1/2$ for $q\geq 1$ and $c_0=1$.
The dynamics of each mode is obeyed by
\begin{eqnarray}
\gamma_q \frac{dX_q(t)}{dt}
&=&
-\frac{\partial {\cal H}_q^{(A)}(X_q)}{\partial X_q} 
+F_q(t)+Z_q(t),
\label{imag_EM}
\end{eqnarray}
where $\gamma_q=\gamma$ is the frictional coefficient in the mode space.
Whereas $q \geq 1$ represents the internal modes, $q=0$ corresponds to a translational mode of a center of mass (indeed, no restoring force $-\partial {\cal H}_0^{(A)}(X_0)/\partial X_0=0$).
The others are converted as $F_q(t)\equiv \sum_{n=0}^N f_n(t) h_{q,n}$ and $Z_q(t)\equiv \sum_{n=0}^N \zeta_n(t) h_{q,n}$ in the same rule. 
The statistics of the Gaussian-distributed noise $Z_q(t)$ is transformed with zero mean $\left< Z_q(t) \right>=0$ and the following covariance:
\begin{eqnarray} 
\left<Z_q(t)Z_{q'}(t')\right>=(2c_q\gamma_qk_BT/N) \delta_{qq'}\delta (t-t').
\end{eqnarray} 
Elastic force is produced by the effective Hamiltonian with a harmonic potential for $q\geq 1$:
\begin{eqnarray}
{\cal H}_q^{(A)}(X_q) = \frac{k_q}{2}X_q^2
\label{Hq_without_fx}
\end{eqnarray}
with $k_q=4k \sin^2{(\pi q/(2N))}$; we adopt $k_q\simeq k(\pi q/N)^2$ expanded around $q/N\ll1$ hereafter (see Appendix for details).
For $q=0$, ${\cal H}^{(A)}_0(X_0)\equiv 0$ is put with $k_{q=0}=0$.
Solving eq.~(\ref{imag_EM})~\cite{JStatMech_Panja_2010,PRE_Sakaue_2012,PRE_Saito_2015} and superimposing the normal modes with eqs.~(\ref{variable_normal}),\,(\ref{kernel_h}), we draw a trajectory for each monomer with a set of noise $\{ Z_q(t')\}$ for $t'\in [0,t]$ as
\begin{eqnarray}
x_n(t) 
&=& \sum_{q=0} \int_0^t dt'\, \frac{F_q(t') +Z_q(t')}{\gamma_q} e^{-(t-t')(k_q/\gamma_q)} h_{q,n}^\dagger
\nonumber \\
&&+\sum_{q=0} X_q(0) e^{-(k_q/\gamma_q)t}h_{q,n}^\dagger.
\label{x_t_Rouse}
\end{eqnarray}
Note that an initial state is in equilibrium, which satisfies the Gaussian distribution dictated by $\left<X_q(0)\right>=0$ and the equipartition of energy $\left< X_q(0)^2 \right>=k_BT/(2Nk_q)$.
\\

{\it --- Generalized Langevin equation ---}
Focusing on the tagged monomer with $x(t)\equiv x_N(t)$, equation~(\ref{x_t_Rouse}) provides an expression with the GLE~\cite{JStatMech_Panja_2010,PRE_Sakaue_2013,PRE_Saito_2015}:
\begin{eqnarray}
\frac{dx(t)}{dt}
&=& \int_0^t ds\, \mu(t-s)f(s) +\eta^{(v)}(t),
\label{GLE_mu}
\end{eqnarray}
where $\mu(t)$ denotes a mobility kernel and $\eta^{(v)}(t)$ is the colored Gaussian-distributed noise with zero mean and covariance $\left< \eta^{(v)} (t)\eta^{(v)} (s) \right>=k_BT\mu(t-s)$.
The kernel qualitatively comprises three components: a center of mass $\mu_{c.m.}(t)$, an instantaneous response $\mu_{ins}(t)$, and an internal configuration $\mu_{\alpha}(t)$ as in
\begin{eqnarray}
\mu (t) = \mu_{c.m.}(t) +\mu_{ins}(t) +\mu_{\alpha}(t),\quad \mu_{c.m.}(t) = \frac{2}{N\gamma}\delta (t),
\nonumber \\
\mu_{ins}(t) \simeq \frac{2}{\gamma}\delta (t), \quad 
\mu_{\alpha}(t) \simeq \frac{1}{\tau_u\gamma} \Biggl| \frac{t}{\tau_u} \Biggr|^{-3/2},
\end{eqnarray}
where $\tau_u\equiv \gamma/k$ or $\tau \equiv \gamma_1/k_1$ denotes the characteristic relaxation time for a monomer or for the entire chain, respectively.
Notably, the normal mode obeyed by eq.~(\ref{imag_EM}) evolves explicitly with the Markov process; however, eq.~(\ref{GLE_mu}) indicates that the GLE picture takes on the colored noise, which lasts until the longest relaxation time $t<\tau$.
The alternation from the Markovian to the non-Markovian processes is a consequence of a loss of a perfect set of state variables that describes the system as Markovian~\cite{StatisticalPhysics_II}.
Despite the change in the noise recognition, however, the equilibrium condition ensures that eq.~(\ref{GLE_mu}) satisfies the fluctuation-dissipation relation (FDR) over all the time frames for $t>0$ (see Appendix): 
\begin{eqnarray}
\left< \Delta x(t) \right> 
=
 \frac{f}{2k_BT} \left< [\delta \Delta x(t)]^2 \right>,
\label{Rouse_FDR}
\end{eqnarray}
where notations for two types of differences are introduced as $\Delta z(t) \equiv z(t)-z(0)$ and $\delta z(t) \equiv z(t)-\left< z(t)\right>$.
Equation~(\ref{Rouse_FDR}) is referred to as an FDR of the first kind.

\subsection{Energy balance}

We can now discuss the energy balance (first law of thermodynamics) based on the Langevin equation.
Recall the thermodynamic limit, which ensures a surface boundary of a system surrounded by a thermal bath, and that fluctuations per unit volume become negligible, as guided by discussion of the surface-to-volume ratio  and the law of large numbers, respectively. 
However, a small system manifests a surface boundary, which may also fluctuate.
In addition to the evident emergence of a surface boundary, there could be multiple choices of surface boundaries; these boundaries govern the number of degrees of freedom eliminated by the projection.
For example, whereas $N$ degrees of freedom are retained in eq.~(\ref{Rouse_em}), the projection enables us to reduce $N$ to a single degree of freedom in the GLE~(\ref{GLE_mu}) for the tagged monomer observation.

Let us observe the difference in energy balance altered by a reduction of the degrees of freedom, which is envisaged as a surface shift:
\begin{eqnarray}
\Delta {\cal H}_f(x(t),x(0)) &=& W(t)-Q(t),
\label{energy_H_balance_tag}
\\
\Delta {\cal H}_f^{(A)}(x(t),x(0)) &=& W^{(A)}(t)-Q^{(A)}(t),
\label{energy_H_balance_full}
\end{eqnarray}
where the effective Hamiltonian is modified by incorporating the term $-f(t) x=-f(t) x_N$ with $x=x_N$ as
\begin{eqnarray}
{\cal H}_f(x,f)&\equiv& {\cal H}(x)-fx = -fx
\label{H_U_W_tag}
\\ 
{\cal H}_f^{(A)}(\{ x_n\},f)&\equiv& {\cal H}^{(A)}(\{ x_n\})-fx_N
\nonumber \\
&=&\frac{1}{2}k\int_0^N dn\,\left(\frac{\partial x_n(t)}{\partial n}\right)^2 -fx_N.
\label{H_U_W}
\end{eqnarray}
The quantities ${\cal H}_f(x,f)$ for the tagged monomer observation or ${\cal H}_f^{(A)}(\{ x_n\},f)$ for the all-monomer observation are distinguished by the superscript $(A)$.

${\cal H}(x)$ is introduced in eq.~(\ref{H_U_W_tag}).
If conservative force $-\partial {\cal H}(x)/\partial x$ acts on the tagged monomer, a part of the integrand in the GLE~(\ref{GLE_mu}) appears as $f(s)-\partial {\cal H}/\partial x|_{x=x(s)}$ instead of just as $f(s)$; however, this conservative force is not present.
Thus, ${\cal H}(x)=const. \equiv 0$, where an indefinite constant in ${\cal H}(x)$ is set to meet with minimum value chosen in eq.~(\ref{Rouse_Hamiltonian}). 
In the second line of eq.~(\ref{H_U_W}), we take a continuum limit representation, where "$\simeq$" is replaced with "$=$" on $\sum_{n=0}^N\simeq \int_0^Ndn$ in notation.

By considering $f(t)$ as an external parameter, we define work as
\begin{eqnarray}
W(t)=W^{(A)}(t)=\int_0^tdt'\, \frac{\partial {\cal H}_f^{(A)}}{\partial f}\biggl|_{t'} \frac{df(t')}{dt'}=-fx(0),
\label{W_tag_all}
\end{eqnarray}
where an infinitesimal quantity $\epsilon$ is incorporated into $\Theta (t+\epsilon)$ in $f(t)\equiv f\Theta (t+\epsilon)$ such that $\Theta (t+\epsilon)=1$ for $t\geq 0$.
From eqs.~(\ref{energy_H_balance_tag}),\,(\ref{energy_H_balance_full}), the heat $Q(t)$ generated along the trajectory for the process from $x(0)$ to $x(t)$ is identified.
Notably, in eqs.~(\ref{energy_H_balance_tag}),\,(\ref{energy_H_balance_full}), the work done on the system or the heat transferred into the thermal bath, respectively, is assigned as positive.

Equation ~(\ref{energy_H_balance_tag})-(\ref{W_tag_all}) leads to one of the main consequences of the surface shift concerning the heat difference:
\begin{eqnarray}
\Delta {\cal H}_f^{(A)}
&=&
Q(t)-Q^{(A)}(t)
\nonumber \\
&=& \sum_{q\geq 1} Nc_q \Delta {\cal H}_q^{(A)}(X_q) (h_{q,N}^\dagger)^2.
\label{U_elastic}
\end{eqnarray}
Thus, if only the tagged monomer is observed, the change in the effective Hamiltonian $\Delta {\cal H}_f^{(A)}$ or the superposition of $\Delta {\cal H}_q^{(A)}(X_q)$ is interpreted as the heat.\footnote{
We can adopt a different definition of work.
Even if the mechanical work (force times displacement) is chosen:
\begin{eqnarray}
W^{(0)}(t)=W^{(0,A)}(t)=f(t)\Delta x(t),
\label{work_tag_all}
\end{eqnarray}
the main statement in the article is not modified.
In fact, although the energy balance is rewritten as
\begin{eqnarray}
\Delta {\cal H}(x(t),x(0)) &=&  W^{(0)}(t)-Q(t),
\\
\Delta {\cal H}^{(A)}(x(t),x(0)) &=& W^{(0,A)}(t)-Q^{(A)}(t), 
\end{eqnarray}
the difference between the observations maintains the same relation irrespective of the work definition:
\begin{eqnarray}
\Delta {\cal H}^{(A)}
-
\Delta {\cal H}
&=&  
\Delta {\cal H}_f^{(A)}
-
\Delta {\cal H}_f
=Q-Q^{(A)},
\label{enr_bal_dif_W0}
\end{eqnarray}
where $\Delta {\cal H}=0$ and $\Delta {\cal H}_f=-f\Delta x$.
For example, either of the work definitions leads to eq.~(\ref{U_elastic}).
}

To make a more convincing argument, we here verify the consistency of eq.~(\ref{U_elastic}) from the viewpoint of heat on the GLE or the mode analyses.
Because heat is defined in the stochastic energetics for the white noise~\cite{SekimotoBook}, we expect that the analogous formalism holds true in the heat for the colored noise.
For an infinitesimal interval $dt$~\cite{JStatMech_Ohkuma_Ohta_2007}, we introduce
\begin{eqnarray}
d' Q(t)
&=& \left[ \int_0^t ds\, \Gamma(t-s)\frac{dx(s)}{ds} -\eta^{(f)}(t) \right] dx(t),
\label{dQ_GLE_Rouse}
\end{eqnarray}
where $\Gamma (t)$ is the frictional kernel and $\eta^{(f)}(t)$ is the colored Gaussian-distributed noise with zero mean and the covariance given by the FDR: $\left< \eta^{(f)}(t)\eta^{(f)}(s) \right>=k_BT \Gamma (t-s)$.
The kernel $\Gamma (t)$ is related to $\mu(t)$ via $\hat{\Gamma}(z)\hat{\mu}(z)=1$ on the Laplace domain, into which $\hat{\phi}(z)=\int_0^\infty dt\ \phi(t) e^{-zt}$ is used with $z$ being the real-space variable. 
The Laplace transformation provides the other form as equivalent to eq.~(\ref{GLE_mu}):
\begin{eqnarray}
f(t)
&=& \int_0^t ds\, \Gamma(t-s)\frac{dx(s)}{ds} -\eta^{(f)}(t).
\label{GLE_Gamma}
\end{eqnarray}
Applying the force balance eq.~(\ref{GLE_Gamma}) to eq.~(\ref{dQ_GLE_Rouse}) and then using $dx(t)=\sum_q dX_q(t)h_{q,N}^\dagger$, we have $d'Q(t)=f(t)dx(t)= \sum_q f(t)dX_q(t)h_{q,N}^\dagger=\sum_q c_qNF_q(t)dX_q(t)(h_{q,N}^\dagger)^2$.
Furthermore, eliminating $F_q(t)$ with eq.~(\ref{imag_EM}), we encounter the mode-space expression:
\begin{eqnarray}
d'Q(t)
&=&
\sum_{q=0}^N c_qN \left[ d'Q^{(A)}_q+d{\cal H}^{(A)}_q \right] (h_{q,N}^\dagger)^2,
\label{Rouse_heat}
\end{eqnarray}
where the first term in the bracket on the right side is defined as 
\begin{eqnarray}
d'Q_q^{(A)}
&\equiv&
\left( \gamma_q \frac{dX_q}{dt} -Z_q(t) \right) \circ dX_q
\label{Rouse_q_heat}
\end{eqnarray}
and the other $d{\cal H}^{(A)}_q$ is from the effective Hamiltonian (eq.~(\ref{Hq_without_fx}) without the $fx$ term).
Stochastic modal motion undergoes the Markov processes; we then explicitly write eq.~(\ref{Rouse_q_heat}) with the Stratonovich multiplication $(\circ)$.
The midpoint definition in the Stratonovich multiplication is appropriate in light of the energy balance~\cite{SekimotoBook} and symmetricity of microscopic reversibility.

We then superimpose $d'Q^{(A)}_q$ or $d{\cal H}^{(A)}_q$, respectively.
Taking the continuum limit in eq.~(\ref{Rouse_Hamiltonian}), 
we convert $\sum_{q=0}^Nc_qN{\cal H}^{(A)}_q(h_{q,N}^\dagger)^2=\int_0^Ndn\,(1/2)k(\partial x_n(t)/\partial n)^2={\cal H}^{(A)}(\{ x_n \})
$ (see Appendix).
In addition, the total heat is counted as $\sum_{q=0}^N c_qN d'Q^{(A)}_q(h_{q,N}^\dagger)^2=\int_0^Ndn\, d'Q_n^{(A)}=d'Q^{(A)}$.
Note that the heat for the $n$-th monomer is defined as
\begin{eqnarray}
d'Q_n^{(A)}
&\equiv&
\left( \gamma \frac{dx_n(t)}{dt} -\zeta_n(t) \right) dx_n(t).
\label{Rouse_n_heat}
\end{eqnarray}
where $\zeta_n(t)\equiv \sum_q Z_q(t)h_{q,n}^\dagger$.
Consequently, $d'Q(t)$ is found to consist of $d'Q^{(A)}$ and $d{\cal H}^{(A)}$, i.e., $d'Q(t)=d'Q^{(A)}+d{\cal H}^{(A)}$.

Thus far, we have shown the energy balance in real space with the variables $\{ x_n(t) \}$, whose dual is equivalent to the normal modes $\{X_q(t)\}$. The question then arises as to whether the energy balance can be introduced in the mode space. 
Let us here define $d'Q_q$ (eq.~(\ref{Rouse_q_heat})) as the $q$-mode components of the heat.
Then, by recalling $d'Q^{(A)}=\sum_{q=0}^N c_qN d'Q^{(A)}_q(h_{q,N}^\dagger)^2$ and comparing eqs.~(\ref{dQ_GLE_Rouse}) with eq.~(\ref{Rouse_heat}) through the eq.~(\ref{imag_EM}) of motion, we find that the elastic conservative force produced by ${\cal H}^{(A)}=\sum_q Nc_q{\cal H}_q$ acts as the random force in the tagged-monomer observation with the GLE.

\section{Self-avoidance and hydrodynamic interactions}
\label{GEq}

Section~\ref{GEq} generalizes the consequences of the Rouse polymer by incorporating the self-avoiding (SA) effect and the hydrodynamic interactions (HIs).
Although the local friction (i.e., $\gamma_q=\gamma$ for the Rouse model) might suggest that eq.~(\ref{Rouse_n_heat}) is trivial in the preceding section, we here introduce the HIs with $\gamma_q$ of eqs.~(\ref{mode_spring_friction}), which provide a useful approximation to treat the heat.
One of the distinctive features of the SA effect or the HIs is the long-range interaction. 
The developed prescription at the larger spatiotemporal scale is to integrate them out as the effective spring constant or the frictional coefficient ~\cite{Doi_Edwards,JCP_Panja_2009,PRE_Sakaue_2013,PRE_Saito_2015}. 
As long as the polymer retains its equilibrium shape under weak perturbations, the same qualitative treatments as those for the Rouse polymer are available by modifying the mode coefficients in eqs.~(\ref{imag_EM}),\,(\ref{Hq_without_fx})~\cite{Doi_Edwards,JCP_Panja_2009,PRE_Sakaue_2013,PRE_Saito_2015}: 
\begin{eqnarray}
k_q =k (q/N)^{2\nu+1}, \quad \gamma_q=\gamma(q/N)^{-\nu(z-2)+1}.
\label{mode_spring_friction}
\end{eqnarray}
where $\nu$ is the Flory exponent.
Although an ideal chain, including the Rouse polymer, takes $\nu=1/2$, the SA interaction may increase the exponents, e.g., $\nu=3/4,\ \simeq 0.588$ in two/three dimensions, respectively.
In addition, $z$ is the dynamical exponent that associates the characteristic relaxation time $\tau$ with the correlation length $R_0$ as $\tau \sim R_0^z$ ($z=3$ for nondraining or $z=2+1/\nu$ for free-draining)~\cite{deGennesBook,Doi_Edwards}\footnote{
Modified exponents to a noninteger in the mode space express the long-range interaction in the real space like a fractional derivative.}.
The independence of modes has been numerically verified as a good approximation~\cite{JCP_Panja_2009}, whereas that for the Rouse polymer is rigorous.
The modified coefficients in eqs.~(\ref{mode_spring_friction}) are just substituted into eqs.~(\ref{variable_normal})-(\ref{Hq_without_fx}) and (\ref{Rouse_heat}); however, the combination of eq.~(\ref{Rouse_heat}) with eq.~(\ref{Rouse_q_heat}) leads to
\begin{eqnarray}
d'Q_n^{(A)}
&=&
\left( \sum_m\gamma_{n-m} \frac{dx_{m}(t)}{dt} -\eta_{n}(t) \right) dx_n(t),
\label{GEq_Q}
\end{eqnarray}
which is substituted for eq.~(\ref{Rouse_n_heat}).
The summation kernel is defined as $\gamma_{n-m} \equiv \sum_{q} \gamma_q h_{q,n}^\dagger h_{q,m}$.
Although the scalings are modified because of the SA effects or the HIs, the qualitative arguments related to the noise or the energetics are unchanged from those for the Rouse model.
Thus, the heat for the SA tagged monomer includes the change in the elastic energy component, i.e., the effective Hamiltonian in the present formalism.

\section{Discussion}
\label{ThermoDy}

In section~\ref{ThermoDy}, we attempt to develop the aforementioned polymer arguments in light of the fluctuation theorem (FT) and stochastic thermodynamics.

\subsection{Fluctuation theorem}

Let us first discuss the FT.
In the preceding section, we studied the heat in the mode space (eq.~(\ref{Rouse_q_heat})), whose definition is given under a Markov process with FDR of the the second kind $\left<Z_q(t)Z_q(t') \right>=(2c_qk_BT\gamma_q/N)\delta(t-t')$.
As in Crooks' FT on the Markov process~\cite{PRE_Crooks_1999}, our starting point is built on the Markov process for the normal modes,
where 
an exponential of the heat divided by the effective thermal energy equates to a ratio between the probability of realizing a forward trajectory given an initial condition and the probability of realizing the reverse trajectory:
\begin{eqnarray}
\exp{\left( \frac{Q_q^{(A)}(t)}{c_qk_BT/N} \right)}
=
\frac{ {\cal P}_q[\{ X_q (\cdot) \}|X_q(0)] }{ {\cal P}_q[\{ X_q^\dagger (\cdot)|X_q^\dagger(0) \}] },
\label{FT_Xq}
\end{eqnarray}
with $Q_q^{(A)}(t)$ or $c_qk_BT/N$ being heat or effective thermal energy, respectively, in the mode space.
The trajectory of the forward process from $X_q (0)$ to $X_q(t)$ with the external parameter $F_q(t')$ (or $f(t')$) for $0\leq t'\leq t$ is denoted by $\{ X_q (\cdot) \}$; also, the reverse trajectory along $X_q^\dagger(t'') \equiv X_q(t-t'')$ with external parameter $F_q^\dagger(t'')=F_q(t-t'')$ for $0\leq t''\leq t$ is denoted by $\{ X_q^\dagger (\cdot) \}$.
${\cal P}_q[\{ X_q (\cdot) \}|X_q(0)]$ or ${\cal P}_q[\{ X^\dagger_q (\cdot)\}|X_q^\dagger(0)]$ represents the probability of the forward or reverse trajectory given the initial position, respectively, on $q$-mode space.
Note that
the external parameters $F_q(t)$ (or $f(t)$) are dropped in the arguments for succinct representation, 
whereas the external parameters conventionally appear together with stochastic observables $X_q(\cdot)$ as a set on the arguments.
The heat in the exponent of eq.~(\ref{FT_Xq}) is obtained by integrating eq.~(\ref{Rouse_q_heat}) along the trajectory $Q_q^{(A)}=\int_{\{ X_q(\cdot) \}}\,\left( \gamma_q dX_q(t')/dt' -Z_q(t') \right) \circ dX_q(t')$.

The FT for a non-Markov process has been investigated analogously~\cite{JStatMech_Zamponi_2005,JStatMech_Ohkuma_Ohta_2007,JStatMech_Aron_2010}.
A direct inspection of the GLE as in ref.~\cite{JStatMech_Ohkuma_Ohta_2007} indicates the FT with the heat definition of eq.~(\ref{dQ_GLE_Rouse}):
\begin{eqnarray}
\exp{\left( \frac{Q(t)}{k_BT} \right)}
&=&
\frac{
{\cal P}[\{x(\cdot)\}|x(0)]
}{
{\cal P}[\{x^\dagger(\cdot)\}|x^\dagger(0)]
},
\label{FT_tag}
\end{eqnarray}
where ${\cal P}[\{x(\cdot)\}|x(0)]$ denotes the probability of realizing a forward trajectory from $x(0)$ to $x(t)$ given an initial position $x(0)$ and ${\cal P}[\{x^\dagger(\cdot)\}|x^\dagger(0)]$ represents the probability of realizing a reverse trajectory.
Recalling the heat relation between the real and the mode space (eq.~(\ref{Rouse_heat})), we encounter
\begin{eqnarray}
\exp{\left( \frac{Q(t)}{k_BT} \right)}
&\equiv&
\exp{\left[ \sum_q (c_qh_{q,N}^\dagger)^2 \frac{Q_q^{(A)} +\Delta {\cal H}_q^{(A)}}{c_qk_BT/N}  \right]}
\nonumber \\
&=&
\prod_q
\left[
\frac{{\cal P}_q[\{X_q (\cdot)\}|X_q(0)]}{{\cal P}_q[\{X_q^\dagger(\cdot)\}|X_q^\dagger(0)]}
\frac{{\cal P}_q(X_q(0))}{{\cal P}_q(X_q(t))}
\right]^{(c_qh_{q,N}^\dagger)^2},
\label{FT_Q_H}
\end{eqnarray}
where eq.~(\ref{FT_Xq}) and the probability density ${\cal P}_q(X_q) ={\cal A}\exp{(-{\cal H}_q^{(A)}/(c_qk_BT/N))}$ on $q$-mode space appear with ${\cal A}$ denoting a normalizing factor.
It is noticeable that the probability density for the initial condition ${\cal P}_q(X_q(0))$ or ${\cal P}_q(X_q(t))$ enters the last equation, which results from a difference in the effective Hamiltonian $\Delta {\cal H}_q$.

We proceed further in $\Delta {\cal H}_q$ from a viewpoint of the colored noise.
The trajectory of $\{x(\cdot)\}$ with a given $x(0)$ is specified by a temporal sequence of $\{ \eta^{(v)}(\cdot) \}$ in eq.~(\ref{GLE_mu}).
As discussed in ref.~\cite{PRE_Saito_Sakaue_2017,PRE_Saito_2017}, when a polymer undergoes anomalous sub- or super-diffusion as the power-law growths $\left< [z(t)-z(0)]^2 \right> \sim t^\alpha$\,($0<\alpha <1$, $\alpha \neq 1$), the stochastic dynamics, which appears to be a non-Markov process in real space, is cast with fractional Brownian motion and decomposed into the modal motion experiencing the Markov process for both the sub-diffusion ($0<\alpha<1$) and the super-diffusion ($1<\alpha<2$).
The present system evolves as sub-diffusion.
Differentiating eq.~(\ref{x_t_Rouse}) with respect to time and extracting the stochastic part, we find that the colored noise in eq.~(\ref{GLE_mu}) is decomposed into
\begin{eqnarray}
\eta^{(v)}(t) 
&=& 
\sum_{q=0} \Biggl[ -\int_0^t dt'\, \frac{k_q}{\gamma_q^2}Z_q(t') e^{-(t-t')(k_q/\gamma_q)} h_{q,N}^\dagger
\nonumber \\
&&
+ \frac{Z_q(t)}{\gamma_q} h_{q,N}^\dagger
- \frac{k_q}{\gamma_q}X_q(0)e^{-(k_q/\gamma_q)t}h_{q,N}^\dagger
 \Biggr].
\label{eta_XqZq}
\end{eqnarray}
Thus, $\{ \eta^{(v)}(\cdot) \}$ is constructed with an additive form of independently distributed Gaussian white noises (a temporal sequence of $\{Z_q(\cdot)h_{q,N}^\dagger \}$ between $0$ and $t$), and the initial position $X_q(0)h_{q,N}^\dagger$ in the mode space.
This implies that the initial conditions appearing in eq.~(\ref{FT_Q_H}) may be interpreted as a consequence of $X_q(0)h_{q,N}^\dagger$ in the colored noise.

\subsection{Stochastic Thermodynamics}

Another remarkable point in the observation difference is solvation of the polymer, that is, the interaction between the polymer and solvent particles.
We here attempt to develop an argument by referring to the notion of a solvated ensemble introduced by Jarzynski in the context of stochastic thermodynamics~\cite{PRX_Jarzynski_2017}.

Although we have thus far traced only the positions of monomers, we here explicitly consider the degrees of freedom of the solvent particles $\{ y_i \}$.
We assume that the total internal energy, including that of the bath, is divided into kinetic energy and potential energy. 
Suppose that the total potential energy exists in the absence of the applied force ($f=0$) and is decomposed into
\begin{eqnarray}
{\cal U}_{tot}(\{ x_n \},\{ y_i \})
&=&
{\cal U}^{(A)}(\{ x_n \}) 
\nonumber \\
&& 
+{\cal U}_{int}(\{ x_n \},\{ y_i \}) +{\cal U}_b(\{ y_i \}).
\label{Hamiltonian_x_p}
\end{eqnarray}
Let the subscripts $n$ and $i$ provide an index for the monomers and the solvent particles, respectively.
The spatial position of the $n$-th monomer or $i$-th solvent particle is assigned as $x_n$ or $y_i$, respectively.
Note that only the $x$-axis is explicitly dealt with in the following discussion.
Indeed, subsequent main consequences are not altered if another spatial dimension
is taken into account.
${\cal U}_{int}$ or ${\cal U}_{b}$ represents the potential energy of the interaction between the polymer and the solvent particles or that of the interaction between solvent particles in the thermal bath, respectively.
The thermal bath is sufficiently large that the effects of the polymer system can be reasonably assumed to be small perturbations.
The potential energy of all of the polymer and solvent particles is denoted by ${\cal U}_{tot}$.

{\it ---solvated ensemble---}

Equivalence of ensemble ensures that the identical macroscopic thermodynamics is obtained in the thermodynamic limit even starting with a microcanonical, canonical, $T{\mathchar`-} P$, or grand canonical ensemble.
The present study is based on the solvated ensemble~\cite{PRX_Jarzynski_2017} under isobaric-isothermal conditions.
Although scaling up a statistical quantity defined at the microscopic level to a macroscopic quantity is not trivial, eq.~(\ref{G_HA}) for the solvated ensemble has been argued to be reduced to the Gibbs free energy at the macroscopic level~\cite{PRX_Jarzynski_2017}.
We now consider the solvated ensemble, whose probability density for the composite system is given with $\beta=1/(k_BT)$ by
\begin{eqnarray}
{\cal P}(\{ x_n \},\{ y_i \}) 
=
\frac{\exp{\left(-\beta [{\cal U}_{tot}+P{\cal V}_b] \right)}}{\int \prod_{n,i} dx_n dy_i\, \exp{\left(-\beta ({\cal U}_{tot}+P{\cal V}_b )\right)}}
\label{P_sol}
\end{eqnarray}
where $P$ is the pressure assumed to be exerted with constant magnitude on the composite system and ${\cal V}_b(\{ y_i \})$ is the fluctuating bulk volume.\footnote{
This argument subtracts the kinetic energy from the outset because it will eventually be eliminated through the momentum integral on average quantities as long as kinetic energy is included as an additive form.
}
The volume may be defined through $P{\cal V}_b(\{ y_i \})=mgh$, where the system plus the bath is in a container closed by a piston and a weight with mass $m$ is placed on the top side under acceleration of gravity $g$.
Here, let $\{ y_i \}$ include the container height $h$.\footnote{
The conventional notation $h$ is used for height, whereas $h_{q,n}$ appears on the transform.
}
When the degrees of freedom $\{ y_i \}$ are integrated out, the probability distribution for the polymer system is given by
\begin{eqnarray}
{\cal P}(\{ x_n \}) 
&=&
e^{-\beta ({\cal H}^{(A)}(\{ x_n \}) -G) },
\label{P_G_HA}
\end{eqnarray}
where the effective Hamiltonian is decomposed into two stochastic quantities: the fluctuating potential energy for the system ${\cal U}^{(A)}(\{ x_n \})$ and the additional fluctuating potential energy $\phi^{(A)}(\{ x_n \})$ satisfying
\begin{eqnarray}
{\cal H}^{(A)}(\{ x_n \}) 
&=&
{\cal U}^{(A)}(\{ x_n \}) +\phi^{(A)}(\{ x_n \}),
\label{H_E_phi_all}
\end{eqnarray}
\begin{eqnarray}
\phi^{(A)}(\{ x_n \})
= -\beta^{-1}
\log{
\frac{\int \prod_{i} dy_i\,e^{-\beta ({\cal U}_{int}+{\cal U}_b+P{\cal V}_b)}}{\int \prod_{i} dy_i\,e^{-\beta ({\cal U}_b+P{\cal V}_b)}}
}.
\label{isoten_all}
\end{eqnarray}
Using eqs.~(\ref{P_G_HA})-(\ref{isoten_all}), the free energy at the small scale and its partition function are defined as
\begin{eqnarray}
G
&=&
-\beta^{-1}\log{{\cal Z}},
\quad 
{\cal Z} =
\int \prod_n dx_n\, e^{-\beta {\cal H}^{(A)}(\{ x_n \}) }.
\label{G_HA}
\end{eqnarray}
The potential of mean force and its spatial derivative yield the mean force averaged over the degrees of freedom of the solvent molecules~\cite{JCP_Kirkwwod_1935,BPC_Roux_Simonson_1999}, and ${\cal H}^{(A)}(\{ x_n \}) $ serves as the potential of mean force.  \footnote{The Hamiltonian of mean force is defined as a general formalism instead of the potential of mean force, as in ref.~\cite{PRX_Jarzynski_2017}.}

We here interpret equations~(\ref{H_E_phi_all}),\,(\ref{isoten_all}),\,(\ref{G_HA}) 
in the sense of the scale separation between the polymer and the solvent particles~\cite{Hill,EPJE_Seifert_2011}.
We then note that ${\cal H}^{(A)}(\{ x_n \})={\cal U}^{(A)}(\{ x_n \}) +\phi^{(A)}(\{ x_n \})$ should correspond to the effective Hamiltonian appearing in the overdamped Langevin equation~(\ref{Rouse_em}).
In terminology, ${\cal H}^{(A)}(\{x_n\})$ is referred to as the Landau free energy at the intermediate scale level in a textbook on stochastic energetics~\cite{SekimotoBook}.
Incidentally, ${\cal H}^{(A)}(\{ x_n \})$ invokes the fluctuating enthalpy at the small scale, such as if the fluctuating volume is defined as ${\cal V}_{b}^{(A)}(\{ x_n \}) \equiv \phi^{(A)}(\{ x_n \})/P$ in the bare representation introduced in ref.~\cite{PRX_Jarzynski_2017}. 
We also note that the excluded volume conventionally utilized in polymer physics is introduced through a different notion rather on the basis of the phenomenological arguments related to two-body interaction~\cite{deGennesBook,Doi_Edwards,Khoklov_Grosberg}.

Upon observation of the tagged monomer, the probability density is given as
\begin{eqnarray}
{\cal P}(x) 
&=&
e^{-\beta ({\cal H}(x) -G) }
\label{P_G_H}
\end{eqnarray}
with fluctuating enthalpy ${\cal H}(x)$\footnote{
The partition function for eqs.~(\ref{P_G_HA}),\,(\ref{P_G_H}) is identical:
\begin{eqnarray}
{\cal Z}
&=&
\prod_n \int dx_n\,e^{-\beta {\cal H}^{(A)}(\{ x_n \})} 
=
\int dx\,e^{-\beta {\cal H}(x)},
\end{eqnarray}
which means that the Gibbs free energy is defined to not be changed from the observation of all the monomers to the observation of the tagged monomer. 
}
\begin{eqnarray}
{\cal H}(x) 
&=&
{\cal U}(x) +\phi(x)
\label{H_E_phi_tag}
\end{eqnarray}
\begin{eqnarray}
\phi(x) 
= 
-\beta^{-1}
\log{
\frac{\int \prod_{\{n\}'} \prod_{i} dx_n dy_i\, e^{-\beta ({\cal U}_{tot}+P{\cal V}_b) }}{\int \prod_{\{n\}'} \prod_{i} dx_n dy_i\, e^{-\beta ({\cal U}_b+P{\cal V}_b) }}
},
\label{isoten_tag}
\end{eqnarray}
where a set of the monomer indices $\{n\}'$ excludes the tagged monomer's index $N$.
The difference alters the recognition of the interaction with the solution as
\begin{eqnarray}
\frac{\partial}{\partial x}
\left[
\phi(x)
-
\left< \phi^{(A)} \right>_{\{x_{n}\}'}
\right]
&=& 
\frac{\partial}{\partial x}\left< {\cal U}^{(A)} \right>_{\{x_{n}\}'},
\label{Dphi_Us}
\end{eqnarray}
where the ensemble average over ${\{x_n\}}'$ is defined as
\begin{eqnarray}
\left< (\cdot) \right>_{\{x_{n}\}'}
&\equiv&
\int \prod_{\{n\}'} dx_n\,
\nonumber \\
&& \times
\left( \cdot \right)
\frac
{\int \prod_{i} dy_i\, e^{-\beta ({\cal U}_{tot}+P{\cal V}_b) }}
{\int \prod_{\{n\}'} \prod_{i} dx_n dy_i\, e^{-\beta ({\cal U}_{tot}+P{\cal V}_b) }}
\label{Prob_x_nneqN}
\end{eqnarray}
with the normalizing condition $\left< 1 \right>_{\{x_{n}\}'} =1$ satisfied.
Equation~(\ref{Dphi_Us}) is one of the manifestations of the solvation due to the dissolved polymer chain except the tagged monomer.
This equation indicates that part of the solvation arises from the chain configurations even in a simple model like a Rouse polymer.

We next consider the other corresponding thermodynamic quantities at the small scale.
Enthalpy and entropy are defined uniquely as state quantities in a macroscopic system; however, according to ref.~\cite{PRX_Jarzynski_2017}, their consistent representation is not unique at the small scale. 
We may have other choices, where each representation has advantages and disadvantages.
The bare representation is one of the consistent formulations, where the enthalpy for a respective observation approach is defined as the state quantities:
\begin{eqnarray}
H^{(A)} &\equiv& \int \prod_n dx_n\, {\cal H}^{(A)}(\{x_n\}){\cal P}(\{ x_n \}),
\\
H &\equiv& \int dx\, {\cal H}(x){\cal P}(x).
\end{eqnarray}
With eq.~(\ref{G_HA}) in mind, the definition of entropy as a state quantity is given by 
\begin{eqnarray}
S^{(A)}
&=&
\frac{H^{(A)} -G}{T},
\qquad 
S
=
\frac{H-G}{T}.
\label{S_def_bare}
\end{eqnarray}
From eqs.~(\ref{S_def_bare}), a change in the entropy is balanced with a change in the enthalpy: 
\begin{eqnarray}
S^{(A)}-S
=
\frac{H^{(A)}-H}{T}.
\label{S_Q_all}
\end{eqnarray}
The bare representation allows us to exploit the Shannon formula ($S^{(A)}= -\int \prod_n dx_n\, {\cal P}(\{ x_n \})\log{{\cal P}(\{ x_n \})}$, and $S=-\int dx\, {\cal P}(x) \log{{\cal P}(x)}$) as an advantage.
Moreover, we consider stochastic entropy~\cite{RepProgPhys_Seifert_2012} given by
\begin{eqnarray}
{\cal S}^{(A)} (\{ x_n(t) \})
&=& -k_B  \log{{\cal P}(\{ x_n(t) \})}
\nonumber \\
{\cal S} (x) &=& -k_B \log{{\cal P}(x(t))}.
\end{eqnarray}
When the detailed balance holds in the equilibrium, the difference in the stochastic entropy is translated into the heat difference:
\begin{eqnarray}
d{\cal S}^{(A)}(\{ x_n(t) \}) -d{\cal S}(x(t))
&=& 
\frac{d{\cal H}^{(A)}(\{ x_n(t) \}) -d{\cal H}(x(t))}{T}
\nonumber \\
&=& \frac{d'Q(t)- d'Q^{(A)}(t) }{T}.
\label{Dif_BareRep}
\end{eqnarray}
We also note that 
the left side of eq.~(\ref{S_def_bare}) is distinguished from 
the entropy defined as $\tilde{S}^{(A)}\equiv -k_B \int \prod_{n,i} {\cal P}(\{ x_n \}, \{ y_i \}) \log{{\cal P}(\{ x_n \}, \{ y_i \})}$~\cite{PRX_Jarzynski_2017} in the partial molar representation, which is one of the consistent formalisms.
While the present polymer formalism looks compatible with the bare representation, the partial molar representation could bring a different advantageous perspective to the polymer system.


\section{Concluding remarks}
\label{conclusion}

The choice of surface boundaries between the system and the thermal bath is inherently involved in the polymer energetics or thermodynamics at the small scale.
We discussed the boundary shift in a polymer resulting from the projection of comparable degrees of freedom in its chainlike structure.
The boundary shift modifies our recognitions of the dynamical characteristics of noises from the white to the colored and also qualitatively alters the interpretation of thermodynamics introduced at the small scale, where the chain elasticity can be considered as heat.
In addition, we have especially focused on the solvation effects, which rely on the boundary choice in the polymer, by utilizing the solvated ensemble in stochastic thermodynamics.

The information obtained from the tagged monomer dynamics is one of the key quantities in experimental cell observations.
Further development of the present approach toward analyses of intracellular dynamics would be interesting.

\section*{Acknowledgement}
The author thanks T. Sakaue for a fruitful discussion and a critical reading.

\section*{Appendix}

\section*{A. Discrete representation}

The restoring force ($q\geq 1$) is diagonalized on the discrete form as
\begin{eqnarray}
&&k\left( x_{n+1}-2x_n+x_{n-1} \right)
\nonumber \\
&=&
k\sum_{q\geq 1} X_q(t)\left( h_{q,n+1}^\dagger-2h_{q,n}^\dagger +h_{q,n-1}^\dagger \right)
\nonumber \\
&=&
-\sum_{q\geq 1} 4k\sin^2{\left(\frac{\pi q}{2N} \right)} X_q(t)\left[\frac{1}{c_q}\cos{\left(\frac{\pi qn}{N} \right)}\right].
\end{eqnarray}
This paper focuses on the asymptotic dynamics obtained by the superposition of the low mode $q/N\ll 1$.
The higher modes do not cause a serious problem because we assume that the microscopic fractal structure persists infinitely downward on the hierarchy, enabling us to renormalize the parameter unit.

\section*{B. FDR}

The solution to eq.~(\ref{imag_EM}) with eq.~(\ref{Hq_without_fx}) is given by
\begin{eqnarray}
X_q(t) 
&=& 
\int_0^t dt'\, \frac{F_q(t') +Z_q(t')}{\gamma_q} e^{-(t-t')(k_q/\gamma_q)} 
\nonumber \\
&&
+ X_q(0) e^{-(k_q/\gamma_q)t}.
\label{Xq_t_Rouse}
\end{eqnarray}
Recall the step force $F_q(t)=(f/N)(-1)^q$ for $t>0$; we then find the FDR for each mode is written as
\begin{eqnarray}
 F_q\left< \Delta X_q(t) \right>
=
F_q^2\frac{\left< \delta \Delta X_q(t)^2 \right>}{2(c_qk_BT/N)}.
\label{mode_2ndFDR}
\end{eqnarray}
We can ensure that the FDR holds on the real space (eq.~(\ref{Rouse_FDR})) using eq.~(\ref{mode_2ndFDR}).
Specifically, the calculation utilizes
$\sum_{q=0}^N c_q F_q\left< \Delta X_q(t) \right>(h_{q,N}^\dagger)^2=(f/N)\left< \Delta x(t) \right>$ and
$\sum_{q=0}^N \left< \delta \Delta X_q(t)^2 \right>(h_{q,N}^\dagger)^2=\left< \delta \Delta x(t)^2 \right>$.

{\it --- Cumulant expansion ---}

Another approach is cumulant expansion.
A characteristic function for the Gaussian distribution is given up to the second moment.
\begin{eqnarray}
\left< \exp{\left[-\frac{f\Delta x(t)}{k_BT}\right]} \right> 
=
\exp{\left[-\frac{f\left< \Delta x(t)\right>}{k_BT} +\frac{f^2\left< \delta \Delta x(t)^2\right>}{2(k_BT)^2)}\right]}  
\label{app_fdx}
\end{eqnarray}
Also, we proceed along a similar line on the mode space.
Because $f\Delta x(t)=\sum_{q} c_qNF_q\Delta X_q(t)(h_{q,N}^\dagger)^2=\sum_{q} (-1)^qNF_q\Delta X_q(t)h_{q,N}^\dagger$, we have
\begin{eqnarray}
\left< \exp{\left[-\frac{f\Delta x(t)}{k_BT}\right]} \right> 
&=&
\left< \exp{\left[-\frac{\sum_{q} (-1)^qF_q\Delta X_q(t)h_{q,N}^\dagger}{(k_BT/N)}\right]} \right>
\nonumber \\
&=&
\prod_q \left< \exp{\left[-\frac{{(-1)^qF_q\Delta X_q(t)h_{q,N}^\dagger}}{k_BT/N}\right]} \right>
\nonumber \\
&=&
\prod_q \exp{}\Biggl[ -\frac{{(-1)^qF_q\left< \Delta X_q(t)\right> h_{q,N}^\dagger}}{k_BT/N} 
\nonumber \\
&&
+\frac{1}{2}\frac{{F_q^2\left< \delta \Delta X_q(t)^2\right> (h_{q,N}^\dagger)^2}}{(k_BT/N)^2}  \Biggr]  
\label{app_FqdXq}
\end{eqnarray}
Given that $h_{q,N}^\dagger=(-1)^q/c_q$ and comparing eqs.~(\ref{app_fdx}),\,(\ref{app_FqdXq}), we obtain
\begin{eqnarray}
&&
-\frac{f\left< \Delta x(t)\right>}{k_BT} +\frac{f^2\left< \delta \Delta x(t)^2\right>}{2(k_BT)^2)}
\\
&=&
\sum_{q=0}^N
\left[
-c_q\frac{F_q\left< \Delta X_q(t) \right>}{k_BT/N}
+
F_q^2
\frac{\left< \delta \Delta X_q(t)^2 \right>}{2(k_BT/N)^2}
\right]\frac{(-1)^q}{c_q}h_{q,N}^\dagger
\nonumber 
\end{eqnarray}
Thus, if eq.~(\ref{mode_2ndFDR}) holds, we arrive at eq.~(\ref{Rouse_FDR}).

\section*{C. Effective elastic energy}

Applying integration by parts, we calculate the heat for infinitesimal interval $dt$:
\begin{eqnarray}
&&
\int_0^N dn\, d'Q_n^{(A)}
\nonumber \\
&=&
\int_0^N dn\, \left( k\frac{\partial^2 x_n(t)}{\partial n^2} +f\Theta (t)\delta (N-\epsilon-n) \right) dx_n(t)
\nonumber\\
&=&
- d \left[ \frac{k}{2} \int_0^N dn\, \frac{\partial x_n(t)}{\partial n} \frac{\partial x_n(t)}{\partial n} \right] 
+f\Theta(t)dx_N(t)
\label{Rouse_micro_enr}
\end{eqnarray}
where the boundary conditions $\partial x_n(t)/\partial n|_{n=0,N}=0$ are imposed and where $\epsilon$ denotes a positive infinitesimal.
The first term in the last line of eq.~(\ref{Rouse_micro_enr}) corresponds to the change in the effective Hamiltonian ${\cal H}^{(A)}(\{ x_n \})$. This term is rephrased as
\begin{eqnarray}
&&
\frac{k}{2} \int_0^N dn\, \frac{\partial x_n(t)}{\partial n} \frac{\partial x_n(t)}{\partial n}  
\nonumber \\
&=&
\left[ \frac{k}{2} \frac{\partial x_n(t)}{\partial n} \frac{\partial^2 x_n(t)}{\partial n^2}  \right]_{n=0}^{n=N}
-\frac{k}{2} \int_0^N dn\, x_n(t) \frac{\partial^2 x_n(t)}{\partial n^2}  
\nonumber \\
&=&
\int_0^N dn\, \sum_{q,q' \geq 1} \frac{1}{2}k \left( \frac{q\pi}{N} \right)^2  X_{q}(t)\circ X_{q'}(t)h_{q,n}^\dagger h_{q',n}^\dagger
\nonumber \\
&=&
2 \sum_{q \geq 1} \frac{1}{2}Nk_qX_q(t)\circ X_q(t),
\label{Elasticity_xn_Xq}
\end{eqnarray}
where we use for $q,q' \geq 1$
\begin{eqnarray}
&&\int_0^N dn\,h_{q,n}^\dagger h_{q',n}^\dagger 
\nonumber \\
&=&
4\int_0^N dn\, \cos{\left( \frac{\pi nq}{N} \right)}\cos{\left( \frac{\pi nq'}{N} \right)}
\nonumber \\
&=&
2\int_0^N dn\, \left[ \cos{\left( \frac{\pi n(q+q')}{N} \right)} +\cos{\left( \frac{\pi n(q-q')}{N} \right)} \right]
\nonumber \\
&=&
\left\{
\begin{array}{ll}
2N & q=q' \\
0 & q\neq q'
\end{array}
\right.
.
\end{eqnarray}

\section*{D. Difference in solvation}

We here show the derivation of eq.~(\ref{Dphi_Us}).
First, the derivative of eq.~(\ref{isoten_all}) with respect to $x_N$ is explicitly written as
\begin{eqnarray}
\frac{\partial \phi^{(A)}(\{ x_n \})}{\partial x_N}
=
\frac
{\int \prod_{i} dy_i\, \frac{\partial {\cal U}_{int}}{\partial x_N} \, e^{-\beta ({\cal U}_{int}+{\cal U}_b+P{\cal V}_b) }}
{\int \prod_{i} dy_i\, e^{-\beta ({\cal U}_{int}+{\cal U}_b+P{\cal V}_b) }}
\end{eqnarray}
Taking an average $\left< (\cdot) \right>_{\{x_{n}\}'}$, we have
\begin{eqnarray}
&&\left< \frac{\partial \phi^{(A)}(\{ x_n \})}{\partial x_N} \right>_{\{x_{n}\}'}
\nonumber \\
&=& 
\int \prod_{\{n\}'} dx_n\,
\frac
{\int \prod_{i}dy_i\, \frac{\partial {\cal U}_{int}}{\partial x_N} \, e^{-\beta ({\cal U}_{int}+{\cal U}_{b}+P{\cal V}_b) }}
{\int \prod_{i} dy_i\, e^{ -\beta ({\cal U}_{int}+{\cal U}_{b}+P{\cal V}_b) }}
\nonumber \\
&& \times
\frac{\int \prod_{i} dy_i\, e^{-\beta ({\cal U}^{(A)}+{\cal U}_{int}+{\cal U}_{b}+P{\cal V}_b) } }
{\int \prod_{\{n\}'} \prod_{i} dx_n dy_i\, e^{-\beta ({\cal U}_{tot}+P{\cal V}_b) }}
\nonumber \\
&=& 
\int \prod_{\{n\}'} dx_n\,
\frac
{\int \prod_{i}dy_i\, \frac{\partial {\cal U}_{int}}{\partial x_N} \, e^{-\beta ({\cal U}_{tot}+P{\cal V}_b) } }
{\int \prod_{\{n\}'} dx_n \prod_{i}  dy_i\, e^{-\beta ({\cal U}_{tot}+P{\cal V}_b) }}.
\label{App_phi_1}
\end{eqnarray}
Next, an explicit expression for $\partial \phi(x)/\partial x$ is written with $x=x_N$ as
\begin{eqnarray}
&&\frac{\partial \phi(x)}{\partial x}
\nonumber \\
&=& 
\frac
{\int \prod_{\{n\}'} dx_n \prod_{i} dy_i\, \left( \frac{\partial {\cal U}^{(A)}}{\partial x_N} +\frac{\partial {\cal U}_{int}}{\partial x_N}\right) \, e^{-\beta ({\cal U}_{tot}+P{\cal V}_b) }}
{\int \prod_{\{n\}'} dx_n \prod_{i} dy_i\, e^{-\beta ({\cal U}_{tot}+P{\cal V}_b)}}
\nonumber\\
\label{App_phi_2}
\end{eqnarray}
Comparing eq.~(\ref{App_phi_1}) with eq.~(\ref{App_phi_2}), we arrive at eq.~(\ref{Dphi_Us}).


\begin{thebibliography}{21}



\bibitem{SekimotoBook}
	{ K. Sekimoto,
  {\it Stochastic Energetics}
  (Springger, Berlin Heidelberg 2010).} 
  
	

\bibitem{PRE_Gelin_Thoss_2009}
	{ M.F. Gelin, and M. Thoss
	Phys. Rev. E {\bf 79}, 051121 (2009).}   

\bibitem{PRL_Seifert_2016}
	{ U. Seifert,
	Phys. Rev. Lett. {\bf 116}, 020601 (2016).}   
	
\bibitem{PRE_Talkner_Hanggi_2016}
	{ P. Talkner, and P. H$\ddot{\rm a}$nggi
	Phys. Rev. E {\bf 94}, 022143 (2016).}  
		
\bibitem{PRX_Jarzynski_2017}
	{ C. Jarzynski,
	Phys. Rev. X {\bf 7}, 011008 (2017).}   	
    







  
\bibitem{deGennesBook}
	{ P.-G.~de~Gennes,
  {\it Scaling Concepts in Polymer Physics}
  (Cornell University Press, Ithaca, 1979).}
  
  
\bibitem{Doi_Edwards}
	{ M. Doi and S. F. Edwards,
  {\it The Theory of Polymer Dynamics}
  (Clarendon, Oxford 1986).}  
  
  

\bibitem{Khoklov_Grosberg}
	{ A.R. Khoklov and A.Y. Grosberg,
  {\it Statistical Physics of Macromolecules}
  (AIP PRESS, New York 1994).}  






\bibitem{StatisticalPhysics_II}
	{ R. Kubo, M. Toda, and N. Hashitsume,
	{\it Statistical Physics II, Nonequilibrium Statistical Physics} 
	(Springer-Verlag Berlin Heidelberg, 1991).}  

\bibitem{PTP_Mori_1965}
  {H. Mori,
	Prog. Theor. Phys., {\bf 33}, 423-455 (1965)}   



\bibitem{JChemPhys_Schiessel_Oshanin_Blumen_1995}
  {H. Schiessel, G. Oshanin, and A. Blumen,
	J. Chem. Phys., 5070 (1995).} 

  
  
  
\bibitem{PRE_Lizana_Barkai_Lomholt_2010}
  {L. Lizana, T. Ambj$\ddot{o}$rnsson, A. Taloni, E. Barkai, and M. A. Lomholt,
	Phys. Rev. E {\bf 81} 051118 (2010).}  

\bibitem{JStatMech_Panja_2010}
  {D. Panja,
	J. Stat. Mech., P06011 (2010).} 	  

\bibitem{PRE_Sakaue_2013}
  {T. Sakaue,
	Phys. Rev. E, {\bf 87}, 040601(R) (2013)}    
  
\bibitem{PRE_Saito_2015}
  {T. Saito and T. Sakaue,
	Phys. Rev. E, {\bf 92}, 012601 (2015)}   	  


    
 


\bibitem{EPL_Harada_2005}
	{ T. Harada,
	Europhys. Lett. {\bf 70}, 49-55 (2005).}   


\bibitem{JStatMech_Zamponi_2005}
	{ F. Zamponi, F. Bonetto, L. F Cugliandolo, and J. Kurchan,
  J. Stat. Mech., P09013 (2005).}   


\bibitem{JStatMech_Ohkuma_Ohta_2007}
	{ T. Ohkuma, and T. Ohta,
  J. Stat. Mech., P10010 (2007).}   
 

\bibitem{JStatMech_Aron_2010}
	{ C. Aron, G. Biroli, and L. F Cugliandolo,
  J. Stat. Mech., P11018 (2010).}   
 


\bibitem{EPJB_Speck_Seifert_2005}
	{ T. Speck and U. Seifert,
  Eur. Phys. J. B {\bf 43}, 521-527 (2005).}  
  

\bibitem{PRE_Dhar_2005}
	{ A. Dhar,
	Phys. Rev. E {\bf 71}, 036126 (2005).}   


\bibitem{PRE_Sharma_Cherayil_2011}
	{ R. Sharma, and B. J. Cherayil,
	Phys. Rev. E {\bf 83}, 041805 (2011).}    
  
  
\bibitem{PRE_Sakaue_2012}
	{ T. Sakaue, T. Saito and H. Wada,
  Phys. Rev. E {\bf 86}, 011804 (2012).}  	 


  
\bibitem{JCP_Panja_2009}
  {D. Panja and G.T. Barkema,
	J. Chem. Phys. {\bf 131}, 154903 (2009).} 	





\bibitem{PRE_Crooks_1999}
	{ G. E. Crooks,
	Phys. Rev. E {\bf 60}, 2721 (1999).} 



\bibitem{PRE_Saito_Sakaue_2017}
	{ T. Saito, and T. Sakaue,
	Phys. Rev. E {\bf 95}, 042143 (2017).} 
	
\bibitem{PRE_Saito_2017}
	{ T. Saito,
	Phys. Rev. E {\bf 96}, 032502 (2017).} 




\bibitem{JCP_Kirkwwod_1935}
	{ J. G. Kirkwood,
	J. Chem. Phys. {\bf 3}, 1 (1935).}  

\bibitem{BPC_Roux_Simonson_1999}
	{ B. Roux, and T. Simonson,
	Biophys. Chem. {\bf 78}, 1 (1999).}   



\bibitem{Hill}
	{ T. L. Hill,
	 {\it Free Energy Transduction and Biochemical Cycle Kinetics}, (Dover, Mineola, New York, 1989).} 



\bibitem{RepProgPhys_Seifert_2012}
	{ U. Seifert,,
	Rep. Prog. Phys. {\bf 75}, 126001 (2012).}   
	 	
	
  	
	
\bibitem{EPJE_Seifert_2011}
	{ U. Seifert,
	Eur. Phys. J. E {\bf 34}, 26 (2011).}   	
	
	

  






	
	

\end{thebibliography}
\end{document}